# Tripling the critical temperature of $KFe_2As_2$ by carrier switch

Jian-Jun Ying[1,2], Ling-Yun Tang[1], Viktor V. Struzhkin[2], Ho-Kwang Mao[2,1], Alexander G. Gavriliuk[3], Ai-Feng Wang[4], Xian-Hui Chen[4] & Xiao-Jia Chen[1,2]

[1]*Center for High Pressure Science and Technology Advanced Research, Shanghai 201203, China*

[2]*Geophysical Laboratory, Carnegie Institution of Washington, Washington, D.C. 20015, USA*

[3]*Institute of Crystallography, Russian Academy of Sciences, Leninsky pr. 59, Moscow 119333, Russia*

[4]*Hefei National Laboratory for Physical Science at Microscale and Department of Physics, University of Science and Technology of China, Hefei, Anhui 230026, China*

**Superconductivity of high critical temperature ($T_c$) superconductors is usually realized through chemical dopant or application of pressure in a similar way to induce charge carriers of either electrons or holes into their parent compounds[1,2]. For chemical doping, superconductivity behaves asymmetrically with the maximum $T_c$ often higher for optimal hole-doping than that of optimal electron-doping on the same parent compound[1,3-5]. However, whether electron carriers could be in favour of higher $T_c$ than holes in such high-$T_c$ superconductors is unknown but attractive. Here we show that the application of pressure can drive $KFe_2As_2$ from hole- to electron-superconductivity after passing the previously reported V-shape or oscillation regime[6-8]. The maximum $T_c$ in the electron-dominated region is tripled to** **the initial value of 3.5 K or the average in the low-pressure hole-dominated region. The structural transition takes place from the tetragonal to collapsed tetragonal phase when the carrier characteristic is changed upon compression. Our results unambiguously offer a new route to further improve superconductivity with huge $T_c$ enhancement for a compound through carrier switch. The strong electronic correlations[9-11] in $KFe_2As_2$ are suggested to account for the unexpected enhancement of superconductivity in the collapsed tetragonal phase.**

$KFe_2As_2$ is considered as the ideal laboratory for investigating the superconducting and normal states of the iron-based superconductors due to the following important facts: (i) Single crystals of this material can be synthesized in extremely high purity. Some external effects such as cation distortion, disorder, and impurities are therefore safely avoided, contributing it as the cleanest sample of iron pnictides. This ensures the investigation of physical properties from the solely intrinsic aspect. (ii) Because $KFe_2As_2$ is located at the end of the overdoped region of $Ba_{1-x}K_xFe_2As_2$ (ref. 3), the concern of the effects on superconductivity from the complicated structural and magnetic transitions, quantum criticality, and electronic nematicity arisen for the underdoped compounds can be removed, which simplifies the study only focusing on the superconducting property itself. (iii) $KFe_2As_2$ possesses the strongest electronic correlations amongst all the iron-based superconductors, classifying it as an electron correlated material[9-11]. This feature might be the origin for its non-disappeared superconductivity even at the ending overdoped position. (iv) Nodal gap symmetry observed in $KFe_2As_2$ is similar to those in cuprates but differs from the nodeless gap structure observed in underdoped iron pnictides[12,13]. Finding a way to tune the structural and physical properties of $KFe_2As_2$ is important not only for understanding all of these interesting features but also for exploring the underlying mechanism of superconductivity for iron-based superconductors.

Maintaining the same chemical composition while applying pressure to change the lattice of a solid has proven to be very effective and powerful in tuning its structural and physical properties in a rather clean way. This method has brought about the superconductivity of many parent iron pnictides including the starting member $BaFe_2As_2$ of hole-doped $Ba_{1-x}K_xFe_2As_2$ and electron-doped $BaFe_{2-x}A_xAs_2$ (A=Ni, Co) systems[3-5]. A comparison study revealed that both the carrier doping through element substitution and the application of pressure changed the lattice in a similar way[2]. Heavy compression on the lattice always drives these superconductors change from the initial tetragonal (T) to collapsed tetragonal (cT) phase[14-17]. The dense cT phase was found to lose superconductivity, which was suggested to result from the suppression of antiferromagnetic spin fluctuations and/or the absence of electronic correlations[18-21]. High-pressure experiments on $KFe_2As_2$ indeed revealed a anomaly of the $T_c$ evolution in either a V- shape or oscillation with pressure, which was explained by the possible change of the pairing symmetry or superconducting gap structure[6-8]. Currently, there has been no report for the existence of the cT phase in $KFe_2As_2$. Examining such a possibility and its relation to superconductivity is highly desired. Interestingly, we indeed observe a cT phase for KFe$_2$As$_2$ at high pressure. To our surprise, a huge enhancement of $T_c$ in the cT phase is found instead of the generally thought destruction of superconductivity[14-17]. This unexpected finding provides a shortcut for synthesizing superconductors with better performance in dense electronic correlated systems.

We have measured the resistivity of a single crystal of KFe$_2$As$_2$ at various pressures up to 30 GPa. Figure 1 shows the temperature dependence of the resistivity at pressures between 12.2 and 16.0 GPa where $T_c$ reaches a maximum of 12 K. The inset presents the enlarged temperature range for the normalized resistivity to 25 K for comparing the superconducting transitions at ambient pressure and 12.9 GPa. $T_c$ is tripled from its initial value of 3.5 K at ambient pressure to the maximum of 12 K at pressure around 12 GPa. All the early measurements revealed that $T_c$ is no more than 4.0 K at ambient pressure or at pressures below 8 GPa (refs. 6-8). The obtained $T_c$ of 12 K is apparently the record for KFe$_2$As$_2$ and three times larger than the ambient value. This is the main finding of our current work.

Figures 2(a) and (b) show the phase diagram in the first run of compression (solid squares) to 30 GPa and the second run including both the decompression (open symbols) and re-compression (solid uptriangles) processes after releasing pressure from 30 GPa. The insets denote the detailed process of changing pressure. Besides the V-shape behaviour below 3 GPa reported previously[6-8], we also observed another dip around 7.5 GPa (Fig. S1 of Supplementary Information (SI), we can not detect superconductivity below 2 K at pressure of 7.3 GPa due to the equipment limitation). The first superconducting phase (SC1) persists up to the critical pressure P1 around 16 GPa and $T_c$ behaves in a wave with an average value of 3 K. The obtained $T_c$'s below 7 GPa are consistent with the previous results[6-8]. Further increasing pressure, another superconducting phase (SC2) suddenly appears, and its $T_c$ decreases with increasing pressure. Figure 2(b) is the phase diagram in the second run when releasing pressure from 30 GPa. The SC2 phase maintains down to the critical pressure P2 around 5 GPa. A typical dome-like behavior of $T_c$ with a maximum of 12 K at around 11 GPa of the SC2 phase was observed. The maximum $T_c$ is 3 times higher than those in the SC1 phase. Below P2, the system is recovered to the beginning SC1 phase. The large difference between P1 and P2 clearly indicates the hysteretic character of the two superconducting phases. This suggests that the first order transition should take place in this compound under pressure. The superconducting transition in the SC2 phase is weak when initially increasing pressure in the first run, while in the second run, it becomes much sharper (SI, Fig. S1). This implies that the SC2 phase is slowly stabilized above P1 in the initial compression run.

$KFe_2As_2$ is classified as the ending compound of the hole-doped $Ba_{1-x}K_xFe_2As_2$ superconductors. Applying pressure on the parent compound $BaFe_2As_2$ has induced superconductivity with $T_c$ variation in a dome-like shape and then destroyed it upon heavy compression[2]. This behavior is very similar to the electron doping in $BaFe_2As_2$ with Ni or Co in the Fe site[4,5]. In both cases, the disappeared superconductivity was considered as the loss of electronic correlations or the suppression of the spin fluctuations[18-21]. If the similar effect is also held for iron-based superconductors, inducing electron carriers should be expected in $KFe_2As_2$ upon compression. This idea has been confirmed for $KFe_2As_2$ from the high-pressure Hall resistivity measurements and thus provides a new understanding for the huge enhancement of superconductivity in this material.

Figure 3 shows the Hall coefficient $R_H$ measured at 20 K at various pressures in both the compression and decompression runs. The $R_H$ value is positive below 16.5 GPa in the compression run but becomes negative after that. This pressure is just the position P1 for separating SC1 and SC2. The opposite signs of $R_H$ signal the different kinds of charge carriers of SC1 and SC2 with a hole character for the former and an electron feature for the latter. The hole character of the SC1 phase is consistent with the initial carrier character because $KFe_2As_2$ has been well identified as a hole heavily overdoped superconductor[22]. The negative value of $R_H$ in the SC2 phase indicates the electron character of its dominative carriers. For the decompression run, $R_H$ maintains negative before a pressure level near P2 but changes to positive after that. This result indicates the electron and hole character for the SC2 and SC1 phase, respectively. Note that the crossover pressures are exactly the same as P1 and P2 which separate two different SC phases. Such significant changes of $R_H$ resemble the $T_c$ variation with pressure in both the compression and decompression runs. It is apparent that the electronic feature of the charge carriers is in favour of high $T_c$ in the SC2 phase compared with the much lower $T_c$ in the hole-doped SC1 phase. The $T_c$ behavior is clearly controlled by the carrier character of $KFe_2As_2$.

The obvious changes of $R_H$ across the pressures P1 and P2 for the compression and decompression runs follow the possible reconstruction of the electronic structure. The higher $T_c$ observed in the SC2 phase is possibly due to the Fermi surface topology change in which the dominative carriers become electrons, thus the doping level and the Fermi surface nesting would be totally different with the initial SC1 phase. This can be examined from the electrical transport properties in the normal states of the two SC phases. We found that all the resistivity data below 30 K can be well fitted by using the formula: $\rho=\rho_0+AT^2$ (Fig. S1 of SI) Figure S2 shows the pressure dependence of the parameter $A$, the black line indicates the initial compression run and the red line indicates the second run of both the decompression and re-compression processes after releasing pressure from 30 GPa. $A$ was found to first decrease with increasing pressure in the SC1 phase but suddenly changes its tendency around P1, at which it gradually increases and then slightly decreases. The sudden change of $A$ reflects the reconstruction of the Fermi surface topology. This behaviour was again observed in the second run in which the critical pressure around P2 for the sudden change of $A$ is much lower than P1 in the compression run. The anomaly of the $A$ behaviour under pressure serves the good detector of the reconstruction of the electronic structure. The consistent pressure levels for P1 and P2 determined from the superconducting transition, Hall coefficient, and transport parameter in the normal state together support the two SC phases with different charge carriers.

Although *A* monotonously increases with decreasing pressure in SC1, *A* shows a dome-like behaviour in SC2 which is similar to the $T_c$ evolution with pressure in this phase. Interestingly, the pressure for the maximum A is coincident with that for the maximum $T_c$ in the SC2 phase. In strongly correlated electron systems, the coefficient *A* of the $T^2$ is often scaled as the strength of the electronic correlations[23,24]. While the complicated behaviour of $T_c$ in the SC1 phase is still under debate[6,7,25], the consistent variation of both $T_c$ and *A* with pressure indicates that the driving force for superconductivity at least in the SC2 phase might be the electronic mechanism.

To obtain more insight into these findings, we examined the structural properties of $KFe_2As_2$ by single-crystal X-ray diffraction at high pressures. Our data indeed confirmed the existence of the first order structural transition of this compound under pressure. In Fig. 4, we show the pressure dependence of the lattice parameters *a* and *c* in a compression run. Although both parameters decrease with increasing pressure before 15 GPa, they behave in a distinct way after that with a clear decrease in c and a gradual increase of *a*. The similar behaviour has been observed in the parent compound $BaFe_2As_2$ and many electron-doped systems[14], in which *a* usually exhibits intermediate anomalous expansion with a *S* shape and *c* suddenly decreases around the critical pressure. The large decrease in *c* is ascribed to As-As hybridization in 122-type compounds[15,16], *i.e.*, the As ion below the top Fe-plane forms a bond with the As ion above the lower Fe-plane, or the onset of $4p_z$ interlayer bond formation[17]. The abrupt change of the lattice parameters has been considered as a signature of the first order structural transition from the T to cT phase.

The axial ratio *c*/*a* exhibits different features across the critical pressure close to P1. In the T phase, *c*/*a* is almost constant at 3.6 below 13 GPa and then rapidly decreases with increasing pressure. Upon further compression, the axial ratio abruptly changes by as much as nearly 33% between the T phase and cT phases. The *c*/*a* ratio of the cT phase maintains a constant of 2.4. The pressure-induced cT phase has been thought to be non-superconducting due to the suppression of the spin fluctuations[18-21]. Contrary to this belief, our results show enhanced superconductivity in the ending compound $KFe_2As_2$ with a $T_c$ three times higher than its initial value or the average value of the SC1 phase. This surprising discovery of such high-$T_c$ might result from the strong electronic correlations in this material. This finding offers a shortcut to study the asymmetry of electron- and hole-doped superconductivity in an iron-based superconductor with the same chemical composition. It also points to a new route to search for higher $T_c$ in iron-based superconductors. The observed cT structure of the SC2 phase also sheds light on the mystery of the second superconducting phase in iron chalcogenides[26].

**METHODS SUMMARY**

**Sample preparation and electrical transport measurements:** High-quality single crystals of $KFe_2As_2$ were grown by conventional solid-state reaction using KAs as self-flux[27]. We took out the high-pressure resistivity and Hall coefficient measurements on these crystals by using the four probe method in the PPMS-9 T (Quantum Design). Single crystals with typical dimensions of 70×70×10 μm$^3$ were loaded in sample chambers made by *c*-BN gasket in diameter of around 130 μm of a miniature diamond anvil cell[28] with 300 μm culets. Four Pt wires were adhered to the sample using the silver epoxy. Daphne oil 7373 was used as a pressure transmitting medium. Pressure was calibrated by using the ruby fluorescence shift at room temperature. Hall coefficient was measured by sweeping the field from -5 T to 5 T at 20 K which is slightly higher than the maximum $T_c$. Hall resistivity is linear with the magnetic field for all the measurements.

**Synchrotron X-ray diffraction**: High-pressure single-crystal diffraction was carried out with the angle-dispersive X-ray diffraction experiments at the synchrotron beam line, sector 13 of the Advanced Phonon Source (APS) of the Argonne National Laboratory. The sample-to-detector distance and the image plate orientation angles were calibrated using LaB$_6$ standard. A monochromatic X-ray beam with a wavelength of 0.3344 Å was used. A symmetric DAC with a pair of 500 μm culet size anvils was used to generate pressure. A stainless steel gasket preindented to 35 microns thick with a 200 μm diameter hole was used as the sample chamber. A small piece of a single crystal with a ruby ball was loaded in the gasket hole. Argon was used as pressure-transmitting medium to ensure better hydrostatic pressure condition. The pressures were also monitored by the ruby fluorescence shifts[29].

**Acknowledgments** This work was supported by EFree, an Energy Frontier Research Center funded by the U.S. Department of Energy (DOE), Office of Science, Office of Basic Energy Sciences (BES) under Award Number DE-SG0001057. The resistivity measurements were supported by the DOE under Grant No. DE-FG02-02ER45955. Use of the APS for XRD measurements was supported by DOE-BES under Contract No. DE-AC02-06CH11357. The sample growth in China was supported by the Natural Science Foundation of China, the "Strategic Priority Research Program (B)" of the Chinese Academy of Sciences, and the National Basic Research Program of China.

**Author Information** Correspondence and requests for materials should be addressed to X.J.C. (xjchen@hpstar.edu.cn).

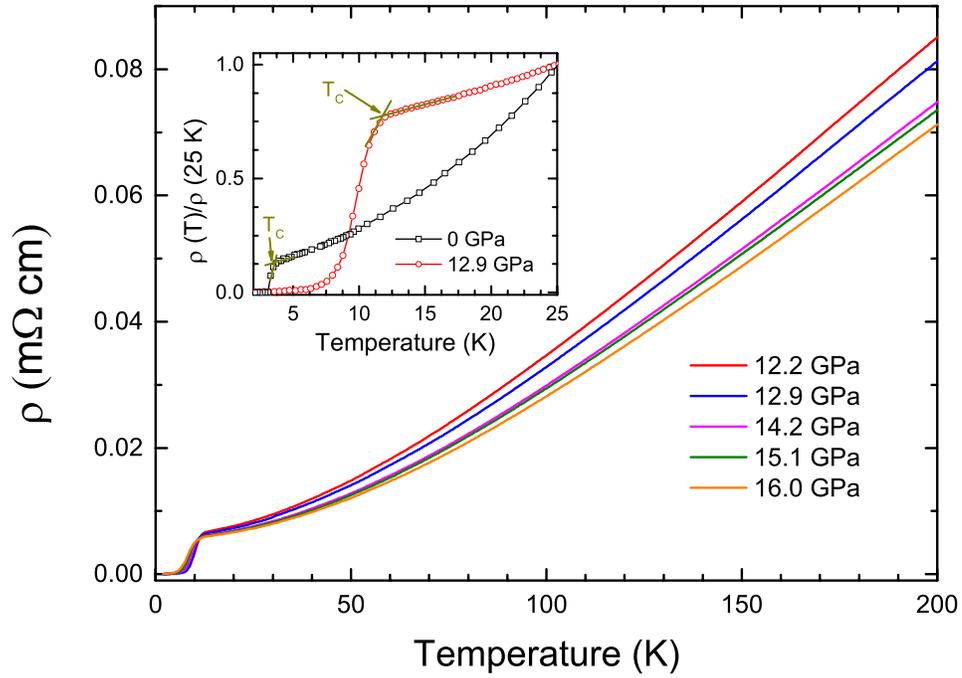

**Figure 1 Temperature dependence of the resistivity of $KFe_2As_2$ at pressures.** The selected resistivity curves are from the measurements upon re-compression between 12.2 and 16.0 GPa when releasing pressure from 30 GPa. Sharp transition and zero resistivity signal the emergence of superconductivity around 12 K. Inset shows the normalized resistivity below 25 K at ambient pressure and 12.9 GPa.

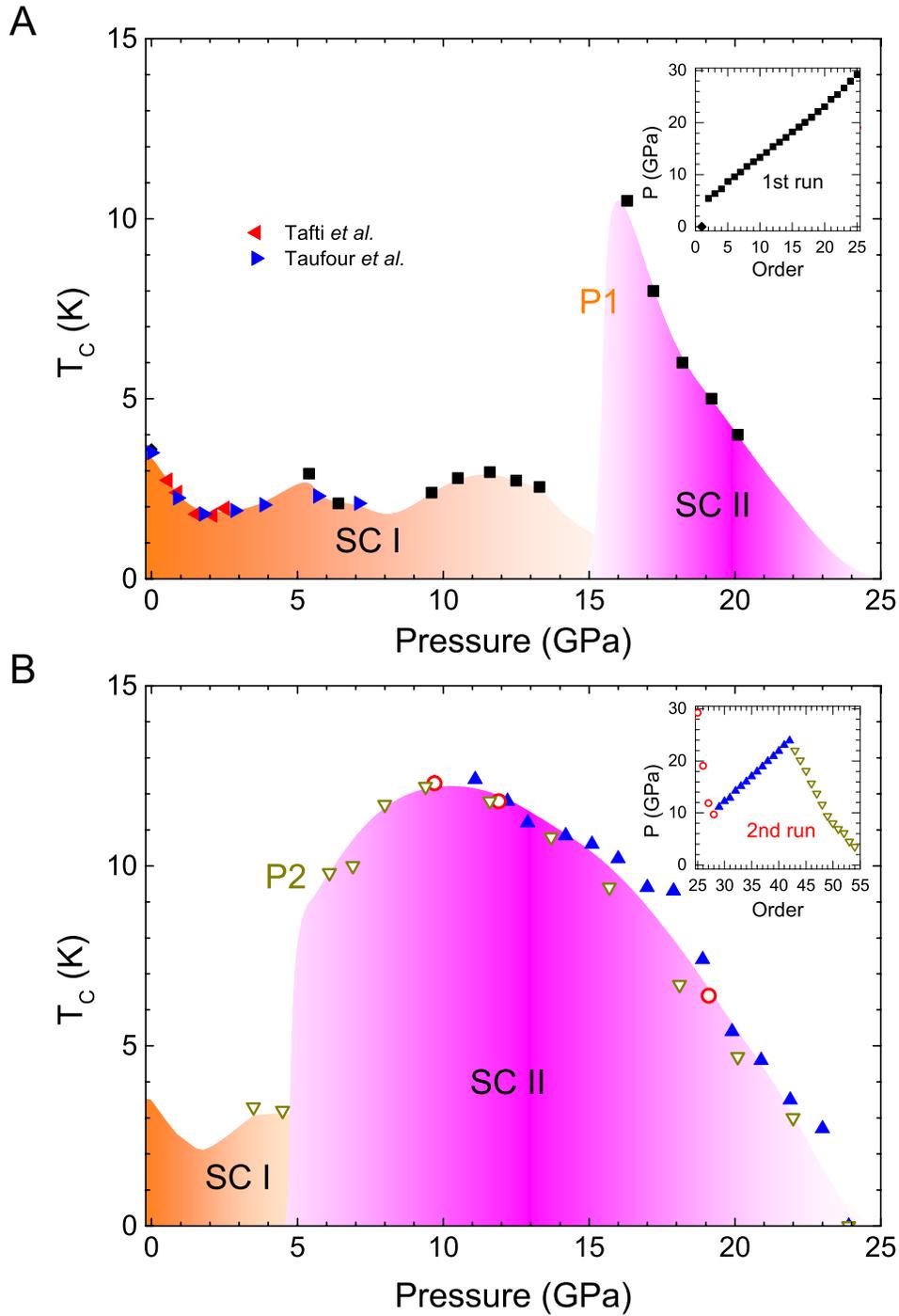

**Figure 2 Phase diagram of KFe$_2$As$_2$ under pressure.** (A) Variation of $T_c$ with pressure in the first run when initially increasing pressure up to 30 GPa. Inset represents the process of applying pressure. Two superconducting phases SCI and SCII are separated at the boundary of P1. The $T_c$ results from the literature are plotted for comparison. (B) $T_c$ behaviour in the second run of changing pressure (details shown in the inset) after releasing from 30 GPa. SC1 and SCII are separated at P2.

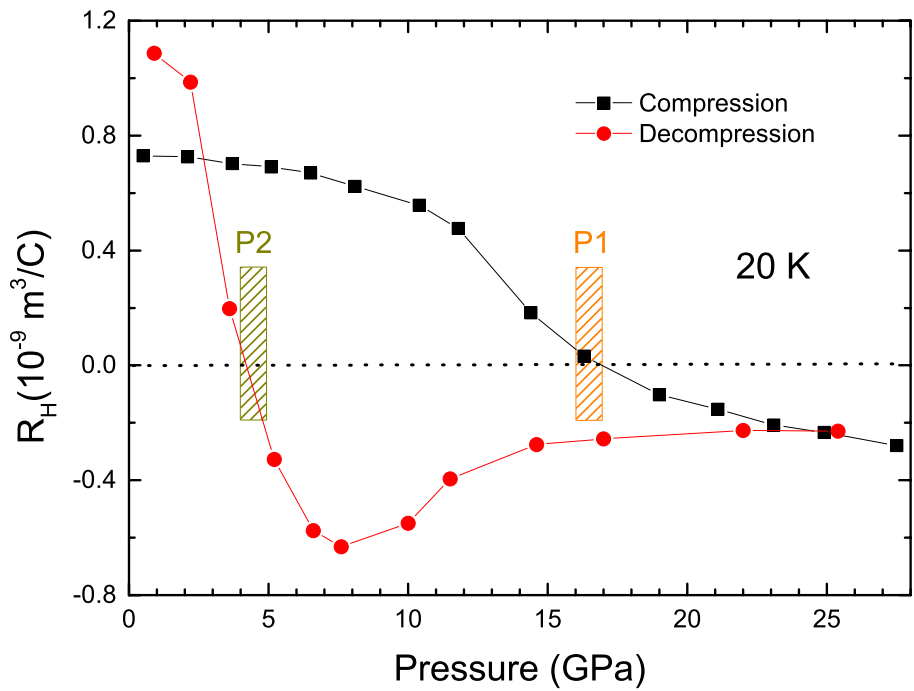

**Figure 3 Hall coefficient of KFe$_2$As$_2$ as a function of pressure.** Positive and negative Hall coefficients correspond to the characteristic of hole and electron carriers of a superconductor. The dashed line at zero denotes the change of the carrier characteristic, from the positive to negative at around P1 in the compression run and the negative to positive at a lower P2 in the decompression run.

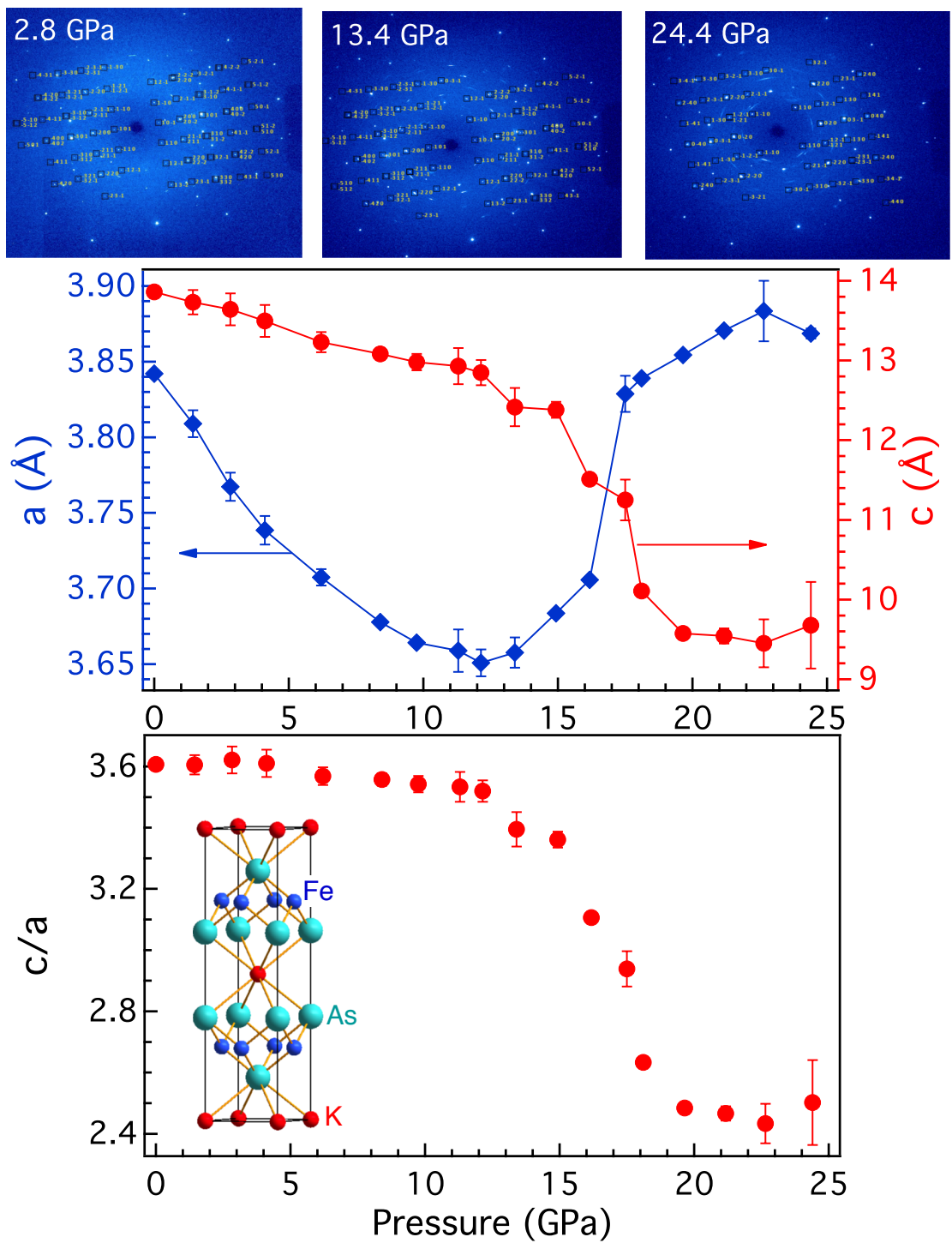

**Figure 4 Structural evolution of KFe$_2$As$_2$ with pressure.** The sudden change of pressure dependence of the lattice parameters *a* and *c*, together with the ratio *c/a* indicates the structural transition from the T to cT phase. The upper panels show single-crystal

synchrotron X-ray diffraction patterns of selected pressures. The marked squares are the refined results to the measured spots of the crystal from which the structural information was deduced.

**Supplementary Information**

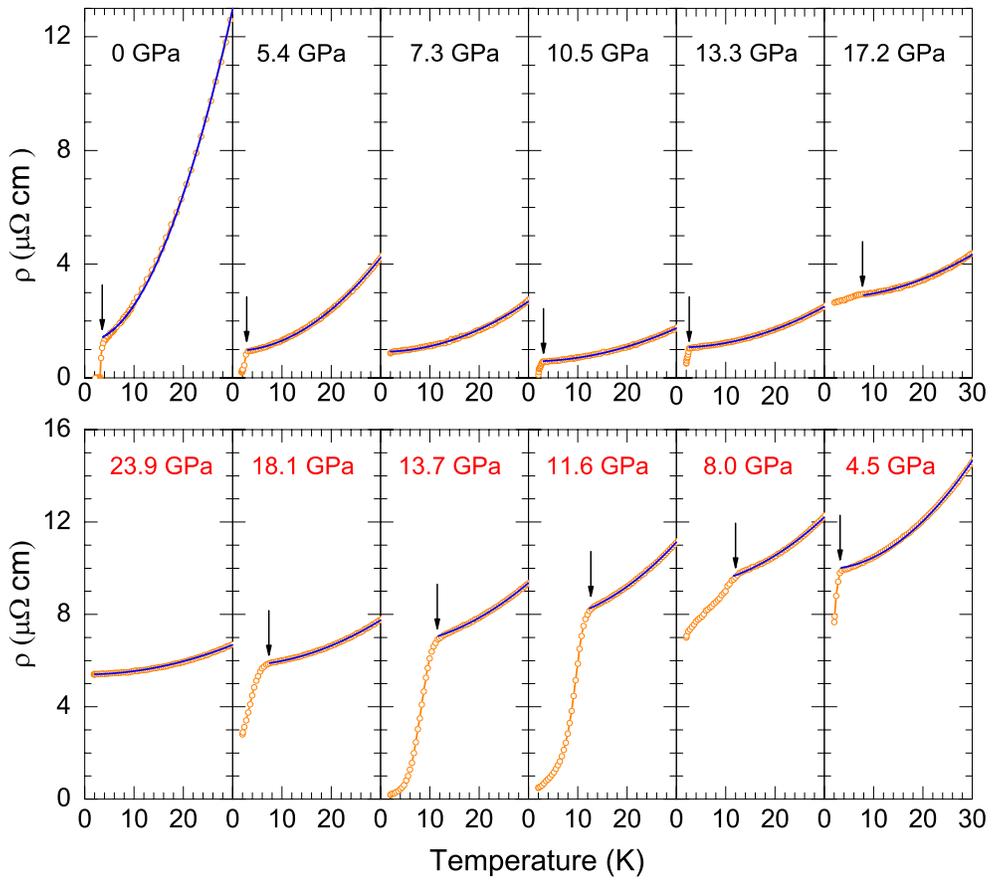

**Figure S1 Normal state resistivity fitting results for $KFe_2As_2$**. Selected resistivity curves around the superconducting transition.. The blue lines are the fitting results below 30 K using the expression $\rho=\rho_0+AT^2$. The upper (lower) panels are the results in the compression (decompression) process.

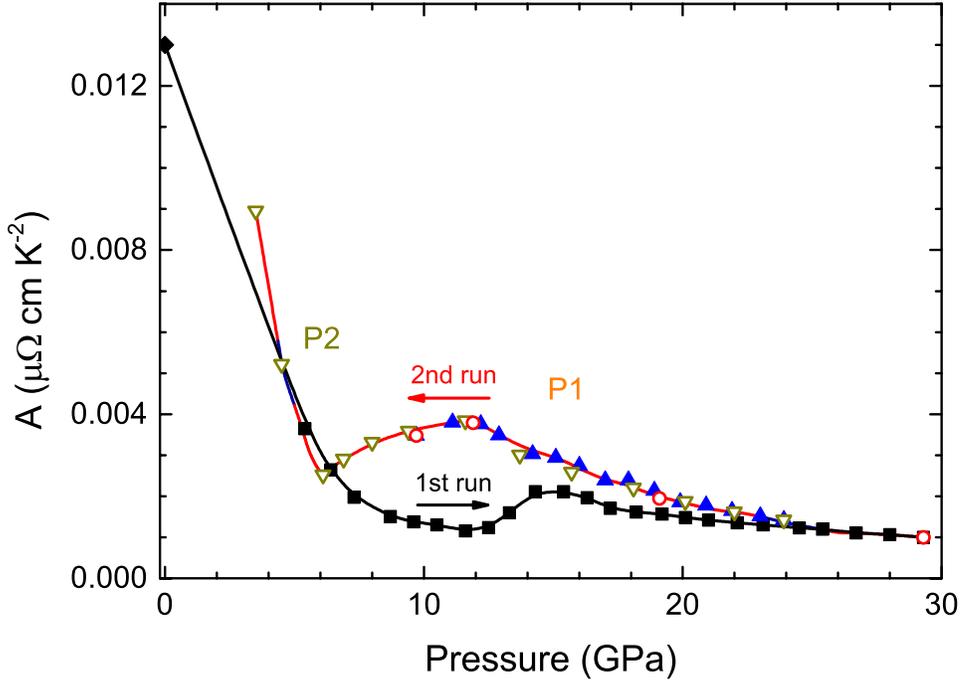

**Figure S2 Pressure dependence of the parameter $A$ for $KFe_2As_2$.** The parameter $A$ is obtained by fitting the normal state resistivity data below 30 K by using the formula: $\rho=\rho_0+AT^2$. The black line indicates the first run of increasing pressure and the red line indicates the second run after releasing pressure from 30 GPa. Around the pressure P1 and P2 at which the two superconducting phases switch each other, $A$ shows anormaly. In the SC2 phase, $A$ shows a dome-like behaviour with a maximum value at the optimal pressure for the occurrence of the maximum $T_c$.